\documentclass{PoS}

\usepackage[utf8]{inputenc}  
\usepackage[T1]{fontenc}       
\usepackage{definitions}
\usepackage{localDefs}
\usepackage{lineno}
\usepackage{caption}           
\usepackage[labelformat=brace]{subcaption} 
\RequirePackage[backend=bibtex,sorting=none,style=numeric-comp]{biblatex} 
\usepackage{FNAL_biblatex}   
\graphicspath{ {./logos/}{./figures/} }

\newlength{\figheight}
\title{Higgs physics at CLIC}

\ShortTitle{Higgs physics at CLIC}

\author{\speaker{Strahinja Lukić}\\
       Vinča Institute, University of Belgrade (RS)\\
       E-mail: \email{strahinja.lukic@cern.ch}}

\author{On behalf of the CLICdp collaboration}

\abstract{The Compact Linear Collider CLIC is an option for a future multi-TeV electron-positron collider, offering the potential for a rich precision physics programme, combined with sensitivity to a wide range of new phenomena. The CLIC physics potential for measurements of the 125~GeV Higgs boson has been studied using full detector simulations for several centre-of-mass energies. The presented results provide crucial input to the energy staging strategy for the CLIC accelerator. The complete physics program for measurements of accessible Higgs boson couplings is presented in this contribution. The ultimate measurement precision is reached when all measurements available at a given centre-of-mass energy are included in combined fits. Operation at a few hundred GeV allows the couplings and width of the Higgs boson to be determined in a model-independent manner through the study of the Higgsstrahlung and WW-fusion processes. At a lepton collider, the measurement of the Higgsstrahlung cross section using the recoil mass technique sets the absolute scale for all Higgs coupling measurements. Recently, it has been shown that including both the leptonic and the hadronic decays of the \PZ boson in this approach improves the statistical precision significantly. Operation at higher centre-of-mass energies provides large statistics for the study of Higgs boson decays and the potential to directly measure the top Yukawa coupling. At the highest centre-of-mass energy (presently assumed to be 3 TeV), the Higgs boson self-coupling can be determined with 10\% precision. }

\FullConference{38th International Conference on High Energy Physics\\
          3-10 August 2016\\
         Chicago, USA}

\begin{document}


\section{Introduction}

The Compact Linear Collider CLIC is based on a novel dual-beam acceleration scheme designed to reach multi-TeV center-of-mass (CM) energy with reasonable cost and size. For this goal, acceleration gradients of 100~MV/m are needed. To keep the RF breakdown rate low, the RF pulse is limited to about 150~ns.
To generate such short RF pulses, a high-intensity high-frequency drive beam accelerated in long pulses is first compressed with delay loops and combiner rings, and then used to generate RF power required to accelerate the physics beam.
The pulse repetition frequency is 50~Hz. The physics beam has a structure with around 300 bunches per pulse, with the bunch spacing of 0.5~ns. The short bunch spacing, alongside with high beam-related backgrounds due to high bunch charge density, require fast readout and excellent time resolution of all detector elements. The CLIC acceleration scheme has been succesfully tested at the CLIC test facility CTF3 at CERN. Accelerating gradients of up to 145~MV/m have been demonstrated.
In the ultimate stage the collider is 50~km long and the physics beams collide with an energy of 3~TeV in the CM.


\subsection{The staged CLIC program}

The CLIC allows to explore a very rich physics program using a staged construction and operation, with three successive stages, each at a higher CM energy. The initial staging scenario described in the CLIC CDR \cite{CLIC_Staging_CDR} was designed from the accelerator standpoint with the ultimate energy stage in mind and comprised three stages with CM energies of 500~GeV, 1.4~TeV and 3~TeV. The Higgs physics studies described in this contribution have been performed with a staging scenario adapted to the physics programme, with the first stage at 350~GeV. Full detector simulation has been performed taking into account hadronic beam-induced backgrounds. More recently the staging scenario has been updated with the first stage at 380~GeV to optimise overall performance, cost and power \cite{clicstagingpaper}. 

\begin{figure}
  \centering
  \includegraphics[width=\sqfigwid]{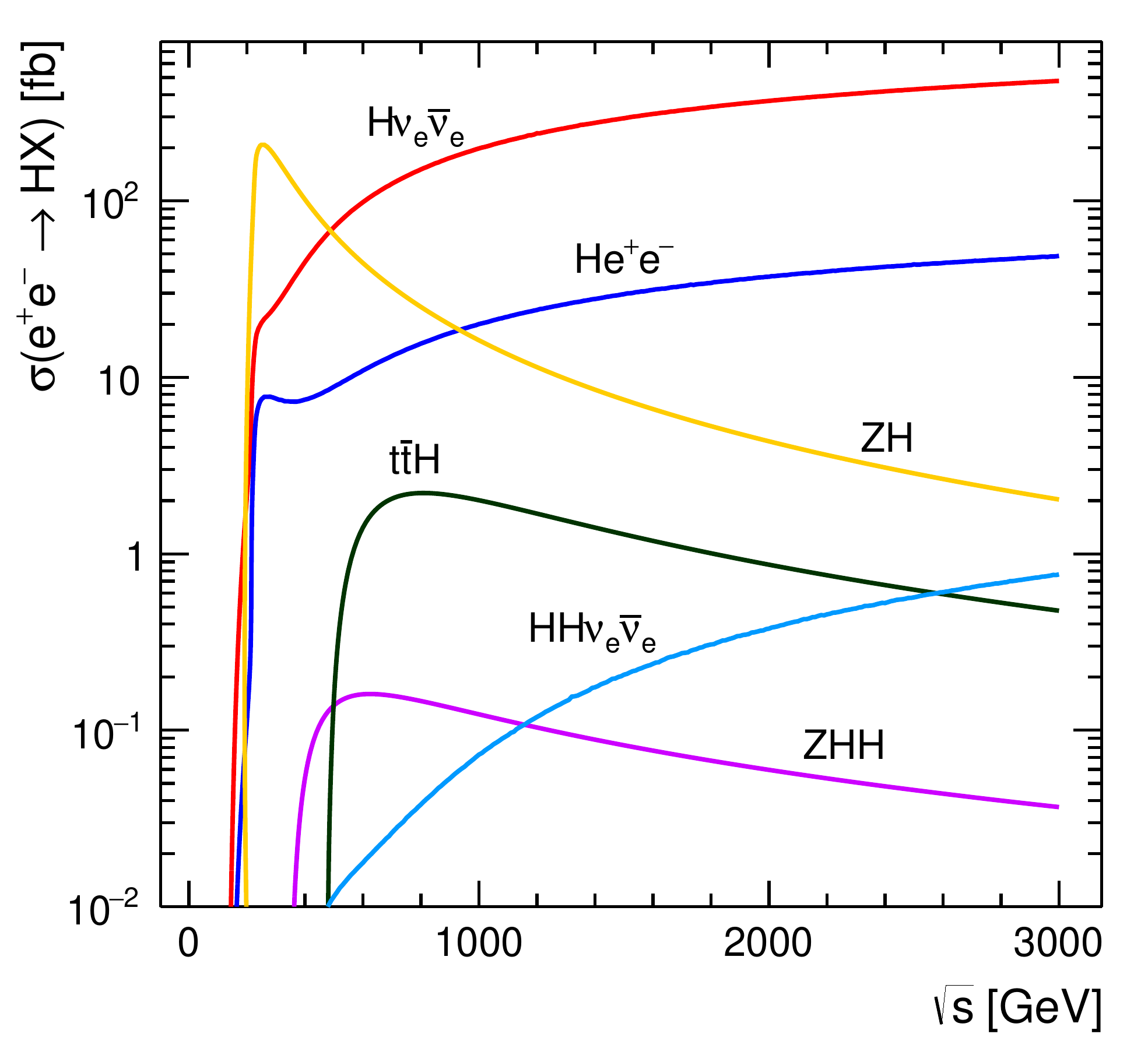}
  \caption{\label{fig:xs} Cross sections for the main Higgs 
         production mechanisms in \epem collisions as a function of CM 
         energy.}
\end{figure}

\figref{fig:xs} shows cross sections for the main Higgs 
production processes in \epem collisions as a function of CM energy. Each of these processes with the exception of the double Higgsstrahlung is accessed individually in the experiment.
By covering the various production processes in three energy stages CLIC is perfectly suited for an in-depth and detailed study of the properties of the Higgs boson. 

The programme of the first CLIC operation stage combines studies of the Higgsstrahlung and \ww fusion processes. The Higgs coupling to the \PZ boson is measured independently of the Higgs decay channels via the Higgsstrahlung recoil analysis. This measurement is complemented by measurements of the Higgs production cross sections times the Higgs decay branching ratios in different production and decay channels to determine the Higgs couplings and the total Higgs decay width in a precise and model-independent way.

At higher energies, the \ww fusion becomes the dominant Higgs production mechanism leading to a production of over a million Higgs bosons during the entire program as indicated in \tabref{tab:xs}. This allows for the measurement of the Higgs couplings with excellent statistical precision and gives access to rare Higgs decays. 

Some of the rarer Higgs production mechanisms are important by their own. 
The Higgs production in conjunction with a pair of top quarks provides a direct measurement of the Top-Yukawa coupling. The measurement of the double Higgs production gives access to the Higgs self coupling and the quartic Higgs coupling to the \PW bosons.

\begin{table}
  \centering
  \caption{ \label{tab:xs}
    Higgs production statistics at each stage with unpolarised beams.
  }
  \setlength{\tabcolsep}{2pt}
  \begin{tabular}{ l r@{\hskip 10pt} r@{\hskip 10pt} r@{\hskip 10pt} r@{\hskip 10pt} r@{\hskip 10pt} r }
    $\roots$ &  \LumiInt   & $N_{\PZ\PH}$ & $N_{\PH\nuenuebar}$ & $N_{\PH\epem}$ & $N_{\ttbar\PH}$ & $N_{\PH\PH\nuenuebar}$ \\ 
    \hline
    350~GeV  & 500\,\fbinv &    68,000    &       17,000        &      3,700 \\
    1.4~TeV  & 1.5\,\abinv &    20,000    &      370,000        &     37,000     &    2,400      &    225   \\
    3~TeV    &  2\,\abinv  &    11,000    &      830,000        &     84,000     &    1,400      &   1200   \\
  \end{tabular}
\end{table}

\section{Higgsstrahlung recoil measurement}

In the Higgsstralung process, a Higgs boson is emitted from a virtual \PZ boson created in the annihilation of a \epem pair. In the recoil measurement only the \PZ-decay is reconstructed and the Higgsstrahlung events are identified by the recoil mass corresponding to the Higgs boson mass. The number of reconstructed HZ events is directly proportional to the Higgsstrahlung cross section $\sigma_{\zhsm}$. 

The decays of the \PZ boson to a pair of electrons or muons have a relatively low BR of 3.5\% per lepton flavour but a clean signature which is clearly separated from the Higgs decay products. 
The decays to a pair of quarks with the total BR of 70\% offer much higher 
statistics but are more 
challenging to reconstruct. The potential confusion with the Higgs decay 
products poses a challenge to ensure the independence of the measurement from the Higgs properties.

\begin{figure*}
  \centering
  \begin{subfigure}{\sqfigwid}
    \includegraphics[width=\sqfigwid]{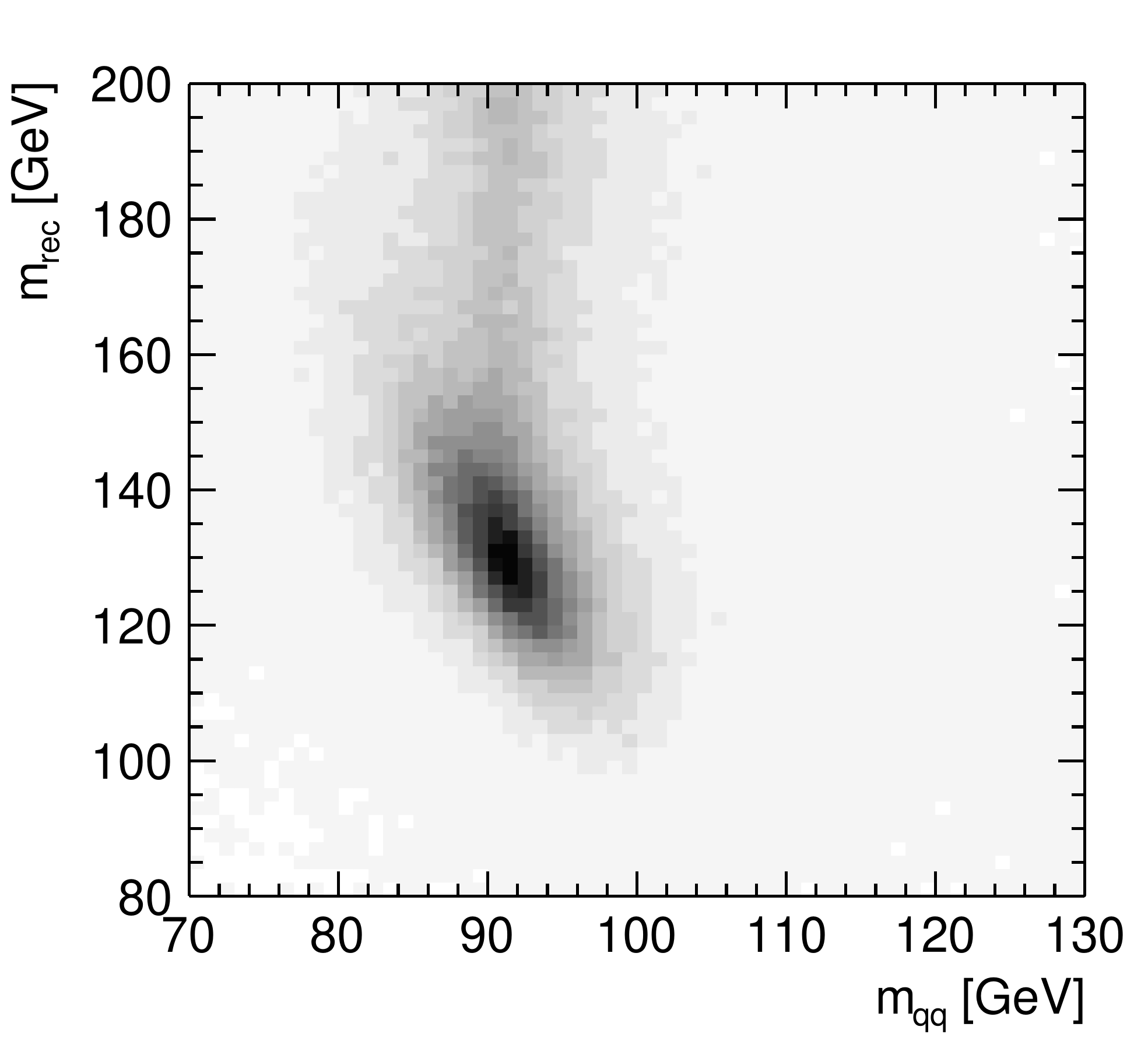}
    \subcaption{\label{fig:mqq_mrec_hzqq}}
  \end{subfigure}
  \begin{subfigure}{\sqfigwid}
    \includegraphics[width=\sqfigwid]{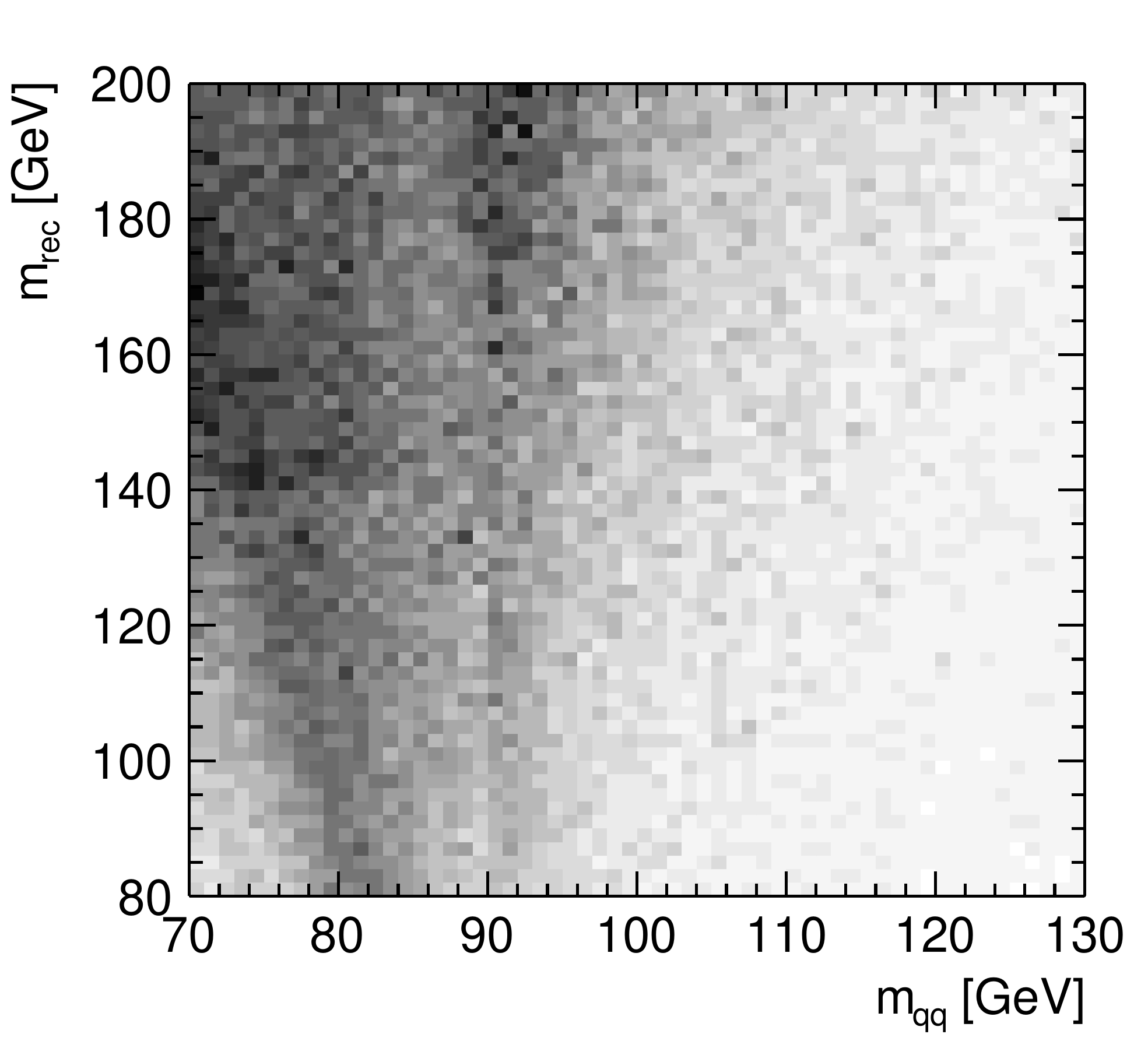}
    \subcaption{\label{fig:mqq_mrec_back}}
  \end{subfigure}
  \caption{\label{fig:hzqq} The 2D distribution of the recoil mass vs. the diquark 
    mass in the analysis of the hadronic \PZ decays in the Higgsstralung process at 350~GeV in CM: 
              \subref{fig:mqq_mrec_hzqq}) signal,
              \subref{fig:mqq_mrec_back}) background.
          }
\end{figure*}

The analysis of the hadronic \PZ
decays is thus tailored specifically to achieve a selection efficiency that 
depends as little as possible on the Higgs decay mode. 
The result of the \zhsm cross section measurement was checked against potential deviations from the Higgs branching ratios predicted in the SM. It was found that any significant bias would require very large deviations which
would anyway be detected at LHC or CLIC \cite{Thomson15}.
\figref{fig:hzqq} shows 
the 2D distribution of the recoil mass vs. the diquark mass - the main 
observable used in the likelihood selection of Higgsstrahlung events in the analysis of the hadronic \PZ decays.

At 350~GeV and 500~\fbinv integrated luminosity the analysis of the $\PZ$ leptonic decay channels results in a statistical accuracy of the $\sigma_{\zhsm}$ measurement of 4.2\%. In comparison, the statistical accuracy in the $\PZ\to\qqbar$ chanel is 1.8\%. The combined precision is 1.65\% 
for $\sigma_{\zhsm}$, or 0.8\% for the Higgs coupling to the Z boson.

\section{$\sigma\times\br{}$ measurements}

The measurement of the observables of the type $\sigma\times\br{}$, where $\sigma$ stands for the Higgs production cross section and $\br{}$ is the branching ratio for the considered decay, includes the Higgsstrahlung, the \ww fusion and the \zz fusion production processes and most of the Standard Model (SM) Higgs decay channels. These measurements complement the Higgsstrahlung 
recoil measurement for the model-independent determination of the Higgs 
couplings and the total decay width.

Among the highlights of this type of measurement is the analysis of hadronic \PZ boson 
decays from Higgsstrahlung events in conjunction with missing energy at 350~GeV. This measurement allows constraining the branching 
ratio for invisible Higgs decays to below 1\%. 
Another highlight is the simultaneous extraction of $\PH\to\bb$, $\PH\to\cc$ and $\PH\to\Pg\Pg$ branching ratios from the Higgsstrahlung and \ww fusion events, thanks to the flavour tagging capabilities at the linear collider detectors.

Above 1~TeV the Higgs samples from \ww fusion are very large and allow measurements with excellent statistical precision. 
The branching ratio of the $\PH\to\bb$ decay is measured with a statistical precision of 0.3\% when data from all three stages are combined. The same analysis allows determination of the Higgs mass with a precision of 33~MeV from the combined data of the 1.4 and 3~TeV stages based on a template fit of the $\bb$ invariant mass distribution. Rare decays, such as $\PH\to\mpmm$ and $\PH\to\gamgam$ with branching ratios of $2\times10^{-4}$ and $2\times10^{-3}$ can also also be measured at the high-energy stages, above 1~TeV.

\section{Rare Higgs production processes}

\subsection{Top Yukawa coupling}

Measurement of the cross section for the Higgs boson production in 
association with two top quarks is directly sensitive to the top 
Yukawa coupling. As a number of important extenstions of the SM predict 
strong deviations of the top Yukawa coupling, this is one of the most 
sensitive probes for the Physics beyond the Standard Model. 

The 1.4 TeV stage of CLIC is particularly suited for this measurement. The cross 
section for the $\epem\to\ttbar\PH$ process is about 60\% of its maximum value at 
around 800 GeV (see \figref{fig:xs}), while the \ttbar background is lower by a factor $\sim3$.
The most sensitive final states are complex, with 6 and 8 jets, 
including 4 $\PQb$ quarks. This requires excellent clustering and flavor-tagging capabilities of the detector.
The measurement precision of the cross section at CLIC is 8.4\%, leading to 4.4\% uncertainty in the top Yukawa coupling \cite{tthRedford}.

\subsection{Trilinear Higgs self coupling}

The Higgs self coupling is accessed via the measurement of the double Higgs production in the \ww fusion. The same final state is produced in a few additional processes involving also the quartic Higgs coupling to the \PW boson. The contribution of the processes not involving the Higgs self coupling is taken into account using generator calculations. 

The cross section for this process is measurable only with high 
energies and luminosities. Using negative \Pem polarization of 80\% the statistical uncertainty of 10\% on the Higgs self coupling is achieved in the 1.4~TeV and 3~TeV stages combined.

\section{The global analysis}

\begin{figure}
  \centering
  \includegraphics[width=\sqfigwid]{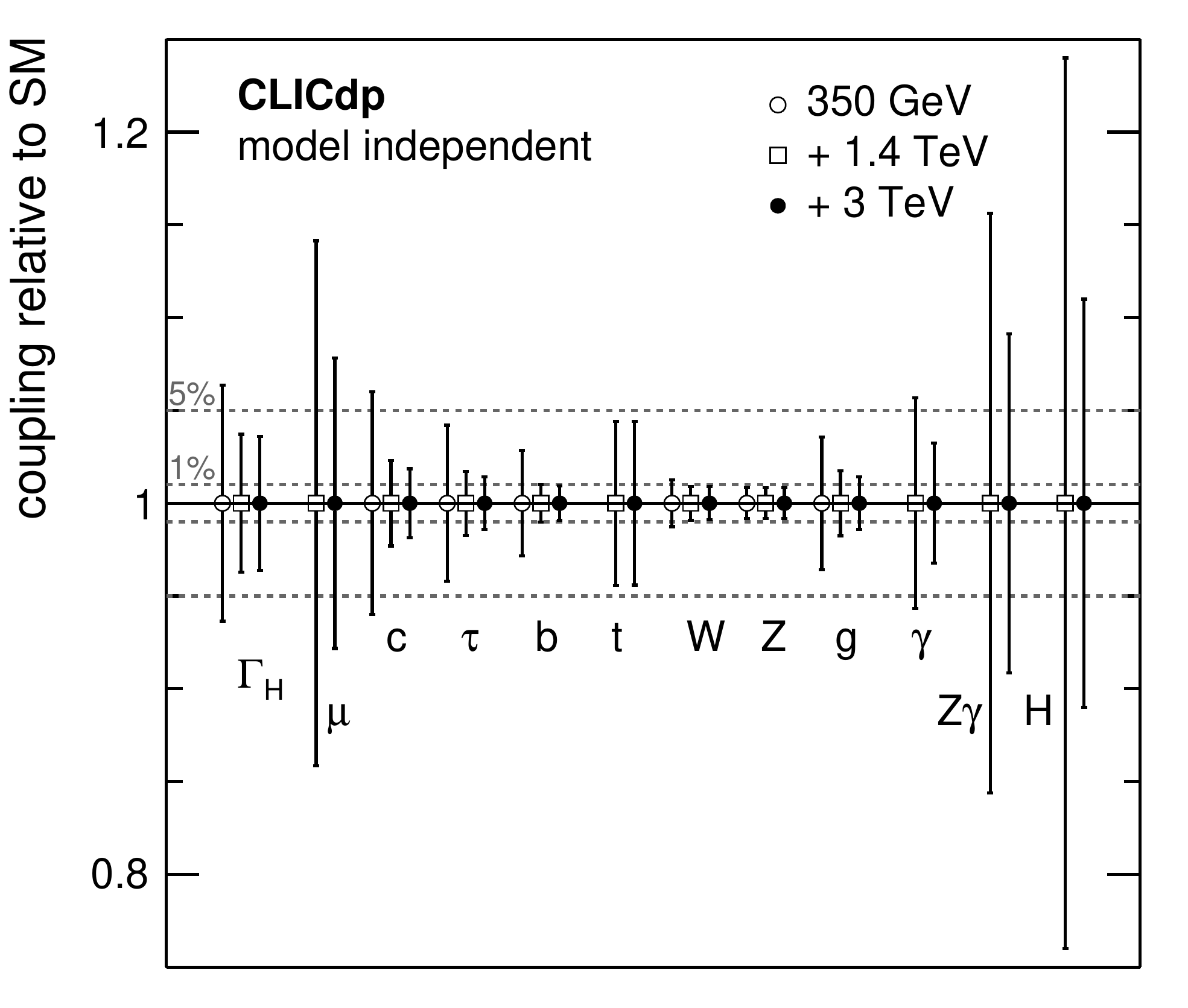}
  \caption{\label{fig:fit} Precision of the model-independent fit 
           of the Higgs couplings and the total decay width using
           data from all three stages of the CLIC programme \cite{clichiggspaper}.}
\end{figure}

The measured Higgsstrahlung cross section and the set of $\sigma\times\br{}$ measurements are used in a global fit to determine the values of the Higgs couplings and the total decay width in a coherent and optimal way. 
For this purpose the $\chi^2$ value,
\begin{equation}
  \label{eq:chi2}
  \chi^2 = \sum_{i} \frac{(C_i/C_i^{\text{measured}} - 1)^2}{\Delta F_i^2} 
\end{equation}
is minimised, where $C_i$ are combinations of coupling constants and 
Higgs width directly corresponding to the measured quantities 
$\sigma\times\br{}$. $\Delta F_i$ are the relative measurement uncertainties of $\sigma\times\br{}$. The total Higgs decay width is extracted with a precision of 3.6\% using data from all three CLIC stages combined. Several of the couplings are determined with a precision below 1\% (see \figref{fig:fit}). 

\section{Conclusions}

Detailed simulation studies of Higgs measurements at CLIC have been 
performed using full detector simulations with overlay of beam 
background and the full reconstruction chain.

CLIC collider offers a very competitive 
Higgs physics program already at the first stage of operation, resulting in a 
Model-independent measurement of the Higgs couplings and of the total 
decay width. Combining the full dataset including 
measurements above 1~TeV, several Higgs couplings are measured to $\mathcal{O}(1\%)$ and several rare Higgs decays with \br{} down to 
$2\times10^{-4}$ are accessed. 
An upper limit of 1\% can be set on the Higgs \br{} for invisible decays. The Higgs mass is determined with 
a statistical precision of 33~MeV from the invariant mass distributions
of the decay products. Finally the top Yukawa coupling and the 
trilinear Higgs self coupling are directly accessed via the measurement 
of the rare Higgs production processes accessible at high energy.

\printbibliography[title=References]


\end{document}